\documentclass[aps,preprint,preprintnumbers,nofootinbib,showpacs]{revtex4-1}
\usepackage{graphicx,color,amsmath}
\begin{document}
\title{Identifying  Glueball  at 3.02 GeV  in Baryonic $B$ Decays}
\author{Y.K. Hsiao$^{1,2}$ and C.Q. Geng$^{1,2,3}$}
\affiliation{
$^{1}$Physics Division, National Center for Theoretical Sciences, Hsinchu, Taiwan 300\\
$^{2}$Department of Physics, National Tsing Hua University, Hsinchu, Taiwan 300
\\
$^3$College of Mathematics \& Physics, Chongqing University of Posts \& Telecommunications, Chongqing, 400065 China}
\date{\today}
\begin{abstract}
We examine the nature of the unknown enhancement around 3 GeV observed by the BABAR collaboration
in the $m_{p\bar p}$ spectrum of the $\bar B^0\to p\bar p D^0$ decay. 
Suspecting that the peak
is a resonance, which can be neither identified as a charmonium state, such as $\eta_c$ or $J/\psi$,
nor classified as one of the light-flavor mesons, we conclude that  it corresponds to a glueball fitted as $X(3020)$ with
$(m_X,\;\Gamma_X)=(3020\pm 8,\; 107\pm 30)\;\text{MeV}$, which could be the first glueball state above 3 GeV.
This state also appears in the $m_{p\bar p}$ spectrum of the $\bar B^0\to p\bar p D^{*0}$ decay.
\end{abstract}

\maketitle
{\it Introduction}---
The glueball ($G$) is a bound state that contains no valence quark but gluons only.
This is because gluons, which are  charged with colors in QCD and  force carriers to
bind quarks becoming mesons and baryons, can also glue themselves together to form a bound state.
Since it is a unique feature purely for the non-Abelian gauge fields,
whether the existence of the gluon condensates can be well established or not
appears to be a real test for QCD.

In principle, the searches for glueballs depend on gluon-rich processes,
such as the radiative $J/\Psi$ decays via $c\bar c\to \gamma gg$.
However, the glueball identifications are inconclusive~\cite{Klempt:2007cp,PofG,status},
which may be illustrated by the following discussions on
the scalar, tensor, and pseudoscalar glueballs.
With the predicted mass around $1.7$ GeV~\cite{LQCD-1,LQCD-2},
the lightest scalar glueball with the quantum number of $J^{PC}=0^{++}$
is allowed to mix with nearby $q\bar q$ mesons in the spectrum.
Since there are two states, $f_0(1500)$ and $f_0(1710)$,
proposed to be composed of the glueball in different mixing scenarios~\cite{mix-0++},
the identification is obscure.
The lightest tensor glueball with $J^{PC}=2^{++}$ is believed to have a mass close to
1.3 GeV in the MIT bag model \cite{2++MBM} and 2.4 GeV in the lattice QCD
calculation~\cite{LQCD-1,LQCD-2}.
For the former, both $f_2(1270)$ and $f'_2(1525)$ as
the ground states of the $2^{++}$ mesons are argued to have
the $2^{++}$ glueball content~\cite{2++1GeV},
while for the later~\cite{PofG},
$f_J(2220)$ ($J=2$ or 4)~\cite{fJ(2220)-1,fJ(2220)-2}
and $f_2(2340)$ \cite{f2(2340)} are considered to be the candidates,
in which  the existence of $f_J(2220)$ is still questionable~\cite{babar-2220}.
Unlike $0^{++}$ and $2^{++}$,
the  difficulty to establish the lightest $0^{-+}$ pseudoscalar glueball is that
the predicted mass around
$2.6$ GeV 
in the lattice QCD calculation~\cite{LQCD-1,LQCD-2}
has no correspondence with the data.
Nonetheless, $\eta(1405)$ seems to be a perfect candidate~\cite{G-1405}.
Particularly,
the unseen  in $\gamma\gamma$ reactions~\cite{gamma-gamma}
reflects that its components are gluons.
In addition,   $X(1835)$, measured first in the $J/\Psi\to \gamma p\bar p$ decays~\cite{gammappbar},
is another possible glueball state~\cite{G-X(1835)} at a mass below 2 GeV.
Interestingly, instead of taking the candidates as the pure glueballs,
the $\eta-\eta'-G$~\cite{eta-eta'-G} and $\eta_c-G$~\cite{eta_c-G}  mixing scenarios
for $\eta(1405)$ and $X(1835)$
are able to allow their own glueball components to be at least 2 GeV, respectively.
Due to the two mixing scenarios, it is  not easy 
to draw a clear conclusion about the glueball state.

Before unfolding the light glueball states,
we may try to explore the heavier ones. Presently,
as the PANDA experiment built to scan heavy glueballs with masses under 5.4 GeV
will not be ready until 2018,
we can only use the decays of the charmonium states,
such as $\eta_c$, $J/\psi$ and $\psi(2S)$, in the mass range of $3.0-3.7$ GeV,
where glueballs with masses around $3$ GeV have been richly predicted.
On the other hand,
although the $B$ meson decays are not regarded as the gluon-rich processes,
they can be more beneficial to offer accesses to a wider detecting range of
heavy glueball productions.
We note that the three-body baryonic 
 decay of $B\to p\bar p M$ with a two-step process $B\to (G\to p\bar p) M$
could be an ideal channel, where $M$ is the recoiled meson.
In particular, one can think of the $G\to p\bar p$ transition
as an inverse process of the $p\bar p$ annihilation, which
has been used at LEAR and PANDA as a gluon-rich process to search for glueballs.
In fact, the process of  $B\to \xi K\to p\bar p K$
has been applied to constrain the narrow resonant state $\xi$,
known as the glueball candidate $f_J(2220)$~\cite{Chua:2002wp,Wang:2005fc}.
Recently,
the BABAR collaboration has observed an unknown enhancement
at $3.0-3.1$ GeV in the $m_{p\bar p}$  spectrum
of $\bar B^0\to p\bar p D^{0}$~\cite{BABAR-2012}.
We shall take that the peak is a sign for a resonant state
as it is unable to be reproduced by the perturbative QCD (pQCD) calculations.
Since the charmonium states, such as $\eta_c$ and $J/\psi$ as well as
the light-flavor mesons are not favored, we 
introduce the glueball state at a mass above 3 GeV as the resonant state.

{\it Data Analysis}---
Before analyzing the unknown peak  
at 3 GeV in the $m_{p\bar p}$  spectrum of $\bar B^0\to p\bar p D^{0}$~\cite{BABAR-2012},
one should emphasize that  the sharp peak around the threshold area of $m_{p\bar p}=(m_p+m_{\bar p})\simeq 2$ GeV is
 commonly observed in $B\to p\bar p M$, which is known as the threshold effect~\cite{HouSoni}.
As this threshold effect dominates the branching ratio,
it may shadow the sign of any new resonance. 
However, in the BABAR's manipulation, 
the threshold effect has been isolated in Fig. 9c of Ref.~\cite{BABAR-2012}
with respect to $m_{D p}>3$ GeV, 
while Fig. 9d of Ref. \cite{BABAR-2012} with respect to $m_{D p}<3$ GeV reveals a resonance
even more obviously. 
As stated by the LHCb collaboration~\cite{LHCb}, 
the $B^-\to p\bar p K^-$ decay is able to offer a clean
environment to study charmonium states and search for  glueballs or exotic states
as $p\bar p$ allows intermediate states of any quantum numbers.
In fact, 
the LHCb in Ref.~\cite{LHCb} has claimed the peaks observed above 2.85 GeV
as resonances, which are further recognized as a serious of charmonium states.
This clearly helps us to find
the true nature of the enhancement at $3.0-3.1$ GeV 
in the $m_{p\bar p}$  spectrum of $\bar B^0\to p\bar p D^{0}$~\cite{BABAR-2012}.
 
\begin{figure}[t!]
\centering
\includegraphics[width=2.5in]{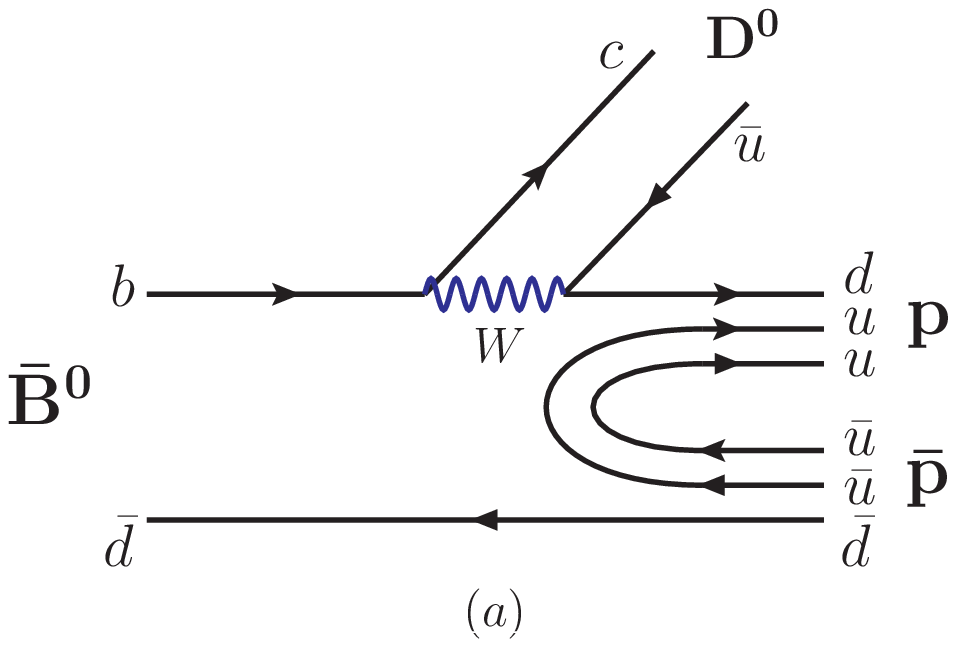}
\includegraphics[width=2.6in]{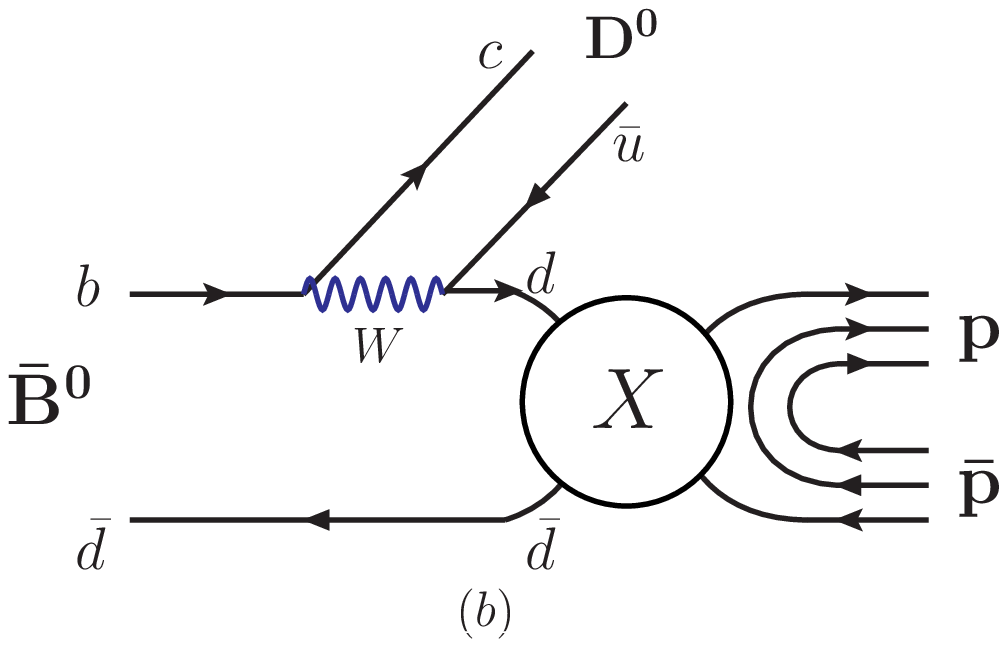}
\caption{The decay of $\bar B^0\to p\bar p D^0$ with the $p\bar p$ productions
by (a) the pQCD effect and (b) the resonance $X$. }\label{figppD}
\end{figure}
In order to explain all data points adopted from Figs. 9c and 9d in Ref. ~\cite{BABAR-2012},
we start with the amplitude based on pQCD counting rules
for $\bar B^0\to p\bar p D^0$ depicted in Fig.~\ref{figppD}a.
The amplitude is given by~\cite{ppD}
\begin{eqnarray}\label{AT}
{\cal A}(\bar B^0\to p\bar p D^{0})&=&\frac{G_F}{\sqrt 2}V_{cb}V_{ud}^*a_2
\langle D^{0}|(\bar c u)_{V-A}|0\rangle\langle p\bar p|(\bar d b)_{V-A}|\bar B^0\rangle\,,
\end{eqnarray}
where $G_F$ is the Fermi constant,  $V_{cb}$ and $V_{ud}$ represent the CKM matrix elements
for the $b\to c\bar u d$ transition at the quark level,
and $(\bar q_1 q_2)_{V(A)}$ stands for $\bar q_1 \gamma_\mu(\gamma_5) q_2$.
For the $D$ meson production, we have
\begin{eqnarray}\label{D0}
\langle D^{0}|(\bar c u)_{V-A}|0\rangle=if_{D} p^\mu\,,
\end{eqnarray}
where $f_D$ is the decay constant of $D$. 
The matrix elements for the $\bar B^0\to p\bar p$ transition  
are parameterized as the most general form~\cite{Btopp}:
\begin{eqnarray}\label{transitionF}
&&\langle p\bar p|\bar d\gamma_\mu b|\bar B^0\rangle=
i\bar u[  g_1\gamma_{\mu}+g_2i\sigma_{\mu\nu}p^\nu +g_3p_{\mu} +g_4 q_\mu +g_5(p_{\bar p}-p_{p})_\mu]\gamma_5v\,,\nonumber\\
&&\langle p\bar p|\bar d\gamma_\mu\gamma_5 b|\bar B^0\rangle=
i\bar u[ f_1\gamma_{\mu}+f_2i\sigma_{\mu\nu}p^\nu +f_3p_{\mu} +f_4 q_\mu +f_5(p_{\bar p}-p_{p})_\mu]        v\,,
\end{eqnarray}
where $p=p_B-p_p-p_{\bar p}$ and $q=p_{p}+p_{\bar p}$ with $p_i\; (i=B,p,\bar{p})$ representing the momenta of the particles. 
The momentum dependences for the form factors $f_j(g_j)$  ($j=1, 2, \cdots, 5$)  
based on pQCD counting rules are~\cite{pQCD}
\begin{eqnarray}\label{figi}
f_j=\frac{D_{f_j}}{t^n}\,,\;\;g_j=\frac{D_{g_j}}{t^n}\,,
\end{eqnarray}
where $t=m_{p\bar p}^2$, $D_{g_1(f_1)}=D_{||}/3\mp 2D_{\overline{||}}/3$,
and $D_{g_k}=-D_{f_k}=-D^k_{||}/3$ ($k=2, 3,\cdots, 5$)
with 
 the reduced constants $D_{||}$, $D_{\overline{||}}$,
and $D^k_{||}$~\cite{derivation}.
By setting $n=3$ to count the number of the hard gluons 
for the $B\to p\bar p$ transition~\cite{pQCD-BB},
the form of $1/t^n$ that peaks at $t\to (m_p+m_{\bar p})^2$
and decreases with increasing $t$
 corresponds with the threshold enhancement.
It is interesting to note that
we have succeeded in explaining
the experimental data observed in baryonic B decays, 
in particular the branching ratios~\cite{radiative,Btopp,ppD,LambdaLambdaKstar} 
of
$B^-\to p\bar p K^{(*)-}(\pi^-)$, $\bar B^0\to p\bar p K^{(*)0}$,
$B^-\to \Lambda \bar p \rho^0(\gamma)$, $\bar B^0\to \Lambda \bar p \pi^+$,
$\bar B^0\to n\bar p D^{*+}$, and $\bar B^0\to p\bar p D^{(*)0}$.
Moreover, the predicted values of ${\cal B}(\bar B\to \Lambda\bar \Lambda \bar K(\pi))$ \cite{LambdaLambdaK}
and ${\cal B}(B^-\to \Lambda\bar p D^{(*)0})$~\cite{ppD}
are approved to agree with the latest 
measurements~\cite{proved}.

\begin{figure}[t!]
\centering
\includegraphics[width=2.05in]{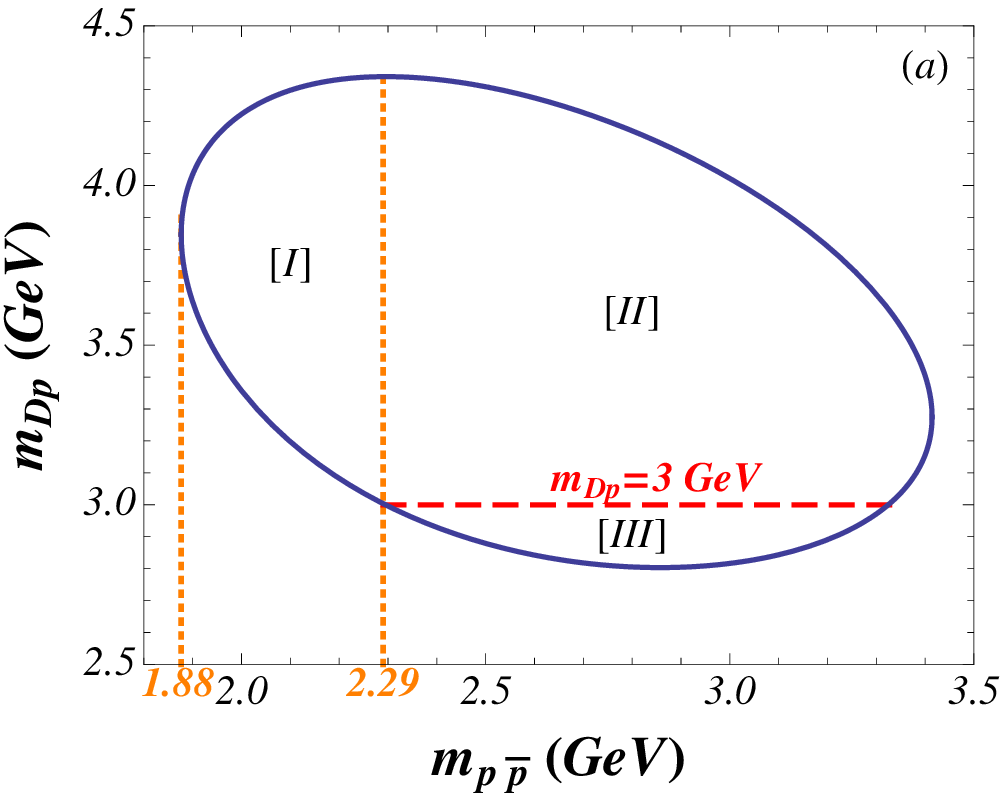}
\includegraphics[width=2in]{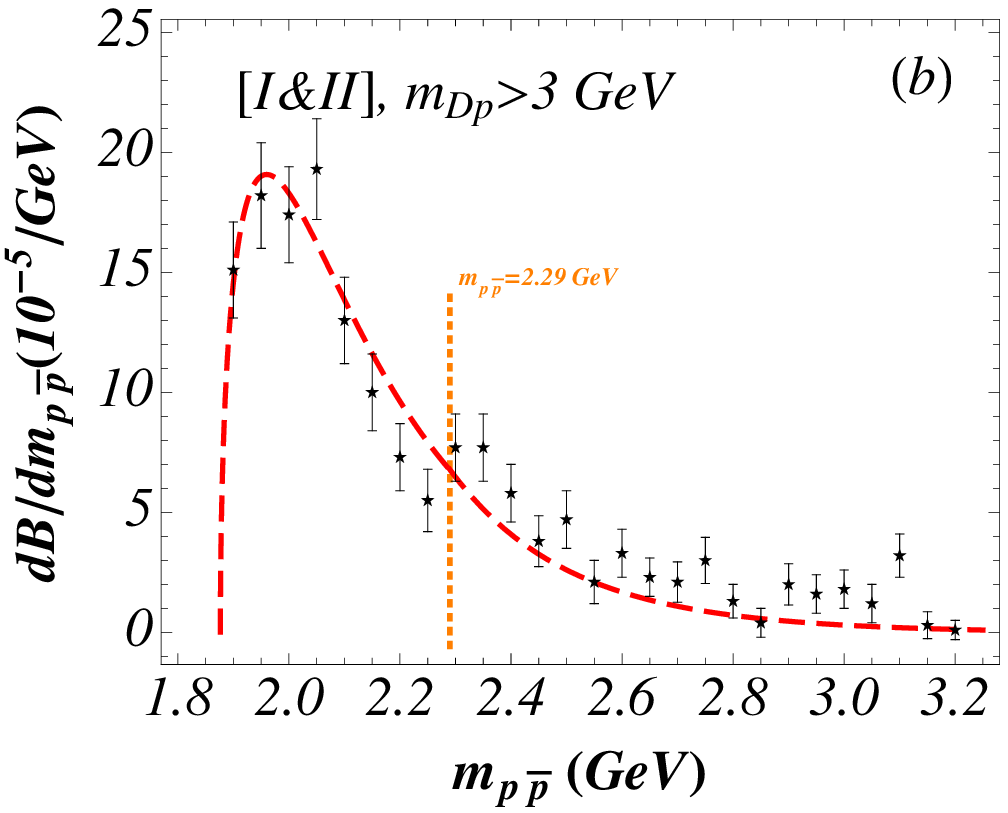}
\includegraphics[width=1.9in]{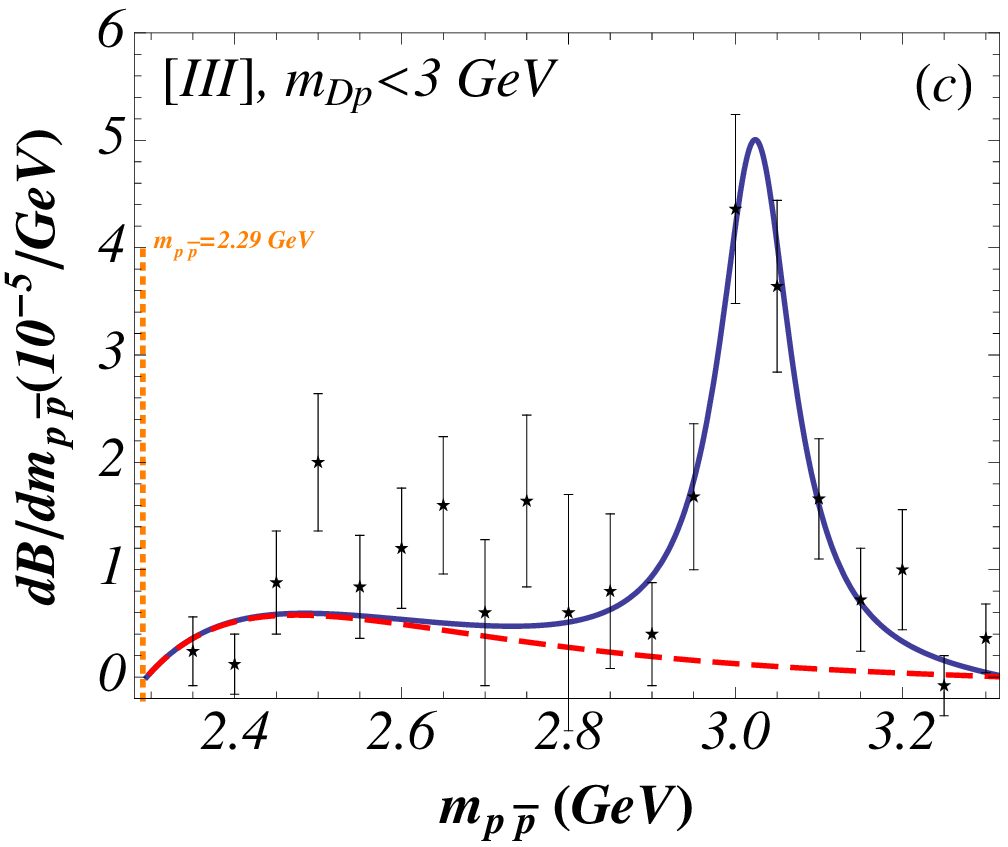}
\caption{
(a) Dalitz plot with the three areas (I, II and III) cut through 
$m_{Dp}=$3 GeV (dashed line) and $m_{p\bar p}=$2.29 GeV (dotted line) in $\bar B^0\to p\bar p D^0$;
 invariant mass spectra as the functions of the invariant mass
$m_{p\bar p}$ with (b) $m_{Dp}>$3 GeV and (c) $m_{Dp}<$3 GeV of the Dalitz plot
in $\bar B^0\to p\bar p D^0$, respectively,
where
the solid line includes the contributions from the resonance and pQCD counting rules
and the dashed lines correspond to those
 without any resonant state, while  the data points are taken from Ref.~\cite{BABAR-2012}.
}\label{diagram}
\end{figure}

In this study, we use the $\chi^2$ fitting 
with the values of $G_F$ , $V_{cb}$, $V_{ud}$, and  $f_D$ from 
Ref.~\cite{pdg}.
We note that the BABAR's manipulation can be realized by cutting
the Dalitz plot of $\bar B^0\to p\bar p D^0$ in Fig.~\ref{diagram}a into 
the three areas (I, II, and III)
by the lines of $m_{D p}=3$ GeV and $m_{p\bar p}=2.29$ GeV.
With the integration of  $m_{D p}>3$ GeV, the area I covers
the data points of the $m_{p\bar p}$ spectrum in Fig. \ref{diagram}b 
starting from 1.88 GeV to 2.29 GeV, including 
the threshold enhancement  isolated in this area,
while the area II corresponds  to 
the data points of $m_{p\bar p}>2.29$ GeV
presenting a limited contribution.
The area III  accords  with
the data points in the $m_{p\bar p}$ spectrum starting from 2.29 GeV
in Fig. \ref{diagram}c, which shows no sign of the threshold effect but with the peak at 3 GeV. 
As seen in Fig.~\ref{diagram}b,  the dashed line in the $m_{p\bar p}$ spectrum
fits well with the data points for the threshold effect 
in the range of $m_{p\bar p}\simeq 2$ GeV
given in Ref.~\cite{BABAR-2012} featured by $f_j(g_j)\propto1/t^3$ in pQCD.
In the fitting, we have $\chi^2/d.o.f=1.9$ with $d.o.f$ denoting the degree of freedom,
which clearly demonstrates the reliability of pQCD counting rules.
In Fig.~\ref{diagram}c, the dashed line in the $m_{p\bar p}$ spectrum
fails to account for the peaking data points. However, 
it fits with the flatness of the non-peaking data points,
which illustrates the suppression above the threshold area.
The fitting leads to $\chi^2/d.o.f=3.95$, 
$2.99$ comes from the 6 points at (2.95, 3.00, 3.05, 3.10, 3.15, 3.20)~GeV, 
showing clearly  the need of a resonant state at 3~GeV.
It seems that raising the dashed line from 0.4 to 1.6~GeV of the height 
in Fig.~\ref{diagram}c can fit the originally unlinked 4 points at (2.50, 2.60, 2.65, 2.75)~GeV, 
resulting in the resonance to be less significant.
Nonetheless, the fits in Figs.~\ref{diagram}b and \ref{diagram}c
depend on the same theoretical inputs, 
which will make the dashed line in Fig.~\ref{diagram}b about 4 times higher too. 
This is obviously unacceptable to the data points, such that
the existence of the resonance at 3~GeV can be established.
 In addition, 
it is interesting to note that the Dalitz plot densities in accordance with the 
areas I, II and III in Fig.~\ref{diagram}a have been measured in Fig.~8a of Ref.~\cite{BABAR-2012}.
It is clear that
 the suppression of the decay rate for the area II 
 also implies the similar smallness for the area III.
Nonetheless, the area III shows a more condense density
converted to be the peak in Fig. \ref{diagram}c, 
which is unable to be traced back to the non-resonant amplitude (dashed line) in Eq. (\ref{AT}). 

We now proceed the second-step identification for the resonance at 3 GeV.
As $\bar B^0\to (M(c\bar c)\to p\bar p)D^0$ is allowed to take place,
with the mass of $M$ around 3 GeV, $J/\psi$ or $\eta_c$ can be the candidate for the resonance.
In Eq.~(\ref{transitionF}),
the $\bar B^0\to p\bar p$ transition is via $\bar B^0(b\bar d)\to (d\bar d\to p\bar p)$.
In pQCD counting rules, one needs three hard gluons for the transition:
one hard gluon is to speed up $\bar d$, while the other two
attach to the valence quarks inside $p\bar p$.
Without being directly related to $p\bar p$ by the hard gluons,
the $d\bar d$ pair can be bounded as the light-flavor meson $M(d\bar d)$.
It is also possible for the $d\bar d$ annihilation, such that  the multi-gluons
are generated to form the glueball $G$ at a mass around 3 GeV.
Therefore, we get three possibilities:
the charmonium $M(c\bar c)$ such as $J/\psi$ and $\eta_c$,
the light-flavor meson $M(d\bar d)$, and the glueball $G$.

Since the dashed line in Fig.~\ref{diagram}c from the pQCD effect has been demonstrated to be small,
we can estimate the resonant contribution to the total branching ratio. 
As a result, we are allowed to test the first possibility of the charmonium   $M(c\bar c)$ as the resonant state at 3 GeV
in terms of a simple relation,  given by
\begin{eqnarray}
{\cal B}(\bar B^0\to (J/\psi\to p\bar p)D^0)\simeq {\cal B}(\bar B^0\to J/\psi D^0){\cal B}(J/\psi\to p\bar p)\,,
\end{eqnarray}
with ${\cal B}(J/\psi\to p\bar p)\simeq 2\times 10^{-3}$ \cite{pdg} as a new  input.
It turns out that ${\cal B}(\bar B^0\to J/\psi D^0)\simeq 4\times 10^{-3}$, which
strongly disagrees with the predicted ${\cal B}(\bar B^0\to J/\psi D^0)$
of order $10^{-6}$~\cite{BtoJpsi-1,BtoJpsi-2}
as well as the experimental upper bound
${\cal B}(\bar B^0\to J/\psi D^0)<1.3\times10^{-5}$~\cite{pdg}.
In addition, it is stated in Ref.~\cite{BABAR-2012} that
the decay width $\Gamma(J/\psi)=93$ keV is not consistent with
the broad 100-200 MeV in the $m_{p\bar p}$ spectrum.
Similarly, we also obtain
${\cal B}(\bar B^0\to \eta_c D^0)\simeq 6.5\times 10^{-3}$, which is much larger than
the predicted ${\cal B}(\bar B^0\to \eta_c D^0)$ of order $10^{-5}$~\cite{BtoJpsi-2}.
Clearly, the resonance cannot be the charmonium.

As seen in  Fig.~\ref{figppD}b for $\bar B^0\to  (X\to p\bar p)D^0$
with X to be $M(d\bar d)$ or $G$,
the relevant amplitude is the same as that in Eq. (\ref{AT}),
while the matrix element of the $\bar B^0\to p\bar p$ transition is given by
\begin{eqnarray}\label{BtoXtopp}
\langle p\bar p|(\bar d b)_{V-A}|\bar B^0\rangle=\langle p\bar p
|X\rangle \frac{i}{(t-m_X^2)+im_X\Gamma_X}\langle X|(\bar d
b)_{V-A}|\bar B^0\rangle\,,
\end{eqnarray}
where $m_X$ and $\Gamma_X$ are the mass and the decay width, respectively.
Consequently, the relevant amplitude of $\bar B^0\to  (X\to p\bar p)D^0$ now reads
\begin{eqnarray}\label{AX}
{\cal A}_R(\bar B^0\to (X\to p\bar p) D^{0})&=&\frac{G_F}{\sqrt 2}V_{cb}V_{ud}^*a_2
\frac{f_D}{(t-m_X^2)+im_X\Gamma_X}\bar u(a+b\gamma_5)v\,,
\end{eqnarray}
with the constants $a$ and $b$.
We note that, no matter what spin the $X$ particle has,
the parameterization for the $\bar B^0\to (X\to p\bar p)$ transition 
can be factored into $a$ and $b$.
Although $a$ and $b$ are in principle energy-dependent,
their values can only be slightly changed with
the deviation for the decay width around 100-200 MeV
compared to the energy range at 3 GeV.
Since the parity determination for the $X$ particle is uncertain,
we set $|a|=|b|$. By taking 
20 data points as our inputs to the combined amplitude
${\cal A}={\cal A}(\bar B^0\to p\bar p D^{0})+{\cal A}_R(\bar B^0\to (X\to p\bar p) D^{0})$,
we fit $|a|=|b|$ and the mass and  decay width of the $X$ particle to be
\begin{eqnarray}\label{mX}
&&|a|=|b|=4.4\pm 1.0\,,\nonumber\\
&&(m_X,\;\Gamma_X)=
(3020\pm 8,\; 107\pm 30)\;\text{MeV}\,,
\end{eqnarray}
respectively. 
Our result with the above resonance is presented as
the solid line in Fig.~\ref{diagram}c. From the figure, we observe that
it can fully explain the peak.
Moreover, compared to $\chi^2/d.o.f\simeq 3.95$ without the resonant amplitude ${\cal A}_R$,
we obtain $\chi^2/d.o.f\simeq 1.17$ to represent 
a good fitting by identifying the peak at 3 GeV as the resonant $X(3020)$.
To fully consider the errors for the fitted mass and decay width of the X resonance,
both the uncertainties from the data points and 
the theoretical inputs~\cite{ppD}
as the background contributions from the pQCD effect
are taken into account, whereas the solid line in Fig.~\ref{diagram}c corresponds to the best fit.
The parameters $|a|$ and $|b|$ fitted to be $4.4\pm 1.0$ can be considered as the size of this process,
 showing the significance to be around 4$\sigma$.
By integrating over $m_{p\bar p}=$2.8-3.2 GeV in the $m_{p\bar p}$ spectrum, 
we give the ratio of the non-resonant and resonant contributions
to be $(6.7^{+3.7}_{-3.0})\%$,
indicating a small background size.
Due to its mass, $X(3020)$
is unlikely to be $M(d\bar d)$. 
In fact, there is no observation
of any light-flavor meson heavier than $f_6(2510)$ in the literature~\cite{pdg},
and the predicted spectrum of the excited mesons
does not span above 2.8 GeV~\cite{Dudek:2011tt}.
This agrees with the study of the hadronic Regge trajectories \cite{Brisudova:1999ut},
where the mass limits are given to be
($2.86\pm 0.11$) and ($3.10\pm 0.11$) GeV
for $n\bar n$ and $s\bar s$ mesons, respectively.
Moreover, the heavier meson with the quark pair inside in the higher state
has more decay channels, resulting in a broader decay width.
Since $f_6(2510)$ has its decay width of $(283\pm 40)$ MeV, it may not be 
possible  for
the heavier $M(d\bar d)$ to shrink  the width back to $(107\pm 30)$ MeV.
As stated in Refs.~\cite{Page:2001gs, Gregory},
the glueball can be ideally observed
in the mass range above 3 GeV, where the productions of the light-flavor mesons
are not able to take place.
As a result, it is reasonable to  recognize $X(3020)$ as the glueball.
Furthermore, it is promising that $X(3020)$ can be one of the glueballs predicted
from various QCD models~\cite{LQCD-1,LQCD-2,G_3GeV,Meyer05,Simonov,Gregory}
in Table~\ref{tab1},
\begin{table}[h!]
\caption{Predicted glueballs around 3 GeV in Refs.~\cite{LQCD-1,LQCD-2,G_3GeV,Meyer05,Simonov,Gregory},
where the units of masses is in MeV.}
\label{tab1}
\begin{tabular}{|c|c|c|}
\hline
$J^{PC}=2^{-+}$                       &$1^{--}$                       &$1^{+-}$                     \\\hline
$3100\pm 30\pm 150$ \cite{LQCD-1}&$3200\pm 200$ \cite{G_3GeV}             &$2940\pm30\pm140$ \cite{LQCD-1} \\
$3040\pm 40\pm 150$ \cite{LQCD-2} &$3240\pm330\pm150$ \cite{Meyer05} &$2980\pm30\pm140$ \cite{LQCD-2} \\
$2950\pm 150$ \cite{G_3GeV}           &$3020$ \cite{Simonov}                         &$3270\pm 340$ \cite{Gregory}\\
\hline
\end{tabular}
\end{table}
where the $2^{-+}$ glueball contains 2 gluons,
while the $1^{--}$ and $1^{+-}$ ones
are allowed to have 3 constituent gluons.
Since $J/\psi(1^{--})$ mainly decays into $ggg$,
the ${\cal O}-J/\psi$ admixture with ${\cal O}$ denoting the $1^{--}$ glueball
is proposed to provide the solution to the so-called $\rho\pi$ puzzle~\cite{rhopi}.
Recently, the experimental data from the charmonium decays at BES and CLEOc
turn out to disfavor this solution~\cite{rhopi-review}.
Nonetheless, one of the original mixing scheme leads to $|m_{\cal O}-m_{J/\psi}|<$ 80 MeV
and $\Gamma_{\cal O}<$ 120 MeV~\cite{vectorO}, agreeing with the fits in Eq.~(\ref{mX}).
Finally,
it is interesting to point out that the same resonance also appears
 in $\bar B^0 \to p\bar p D^{*0}$~\cite{BABAR-2012}.
The combination of the two sets of data should be statistically more convincing.

{\it Discussions and Conclusions}---
We remark that,  via $d\bar d$,
the resonance at 3 GeV can be also explained by a bound state, such as 
the excited $N^*\bar N^*$ bound state with $N^*$ 
being one of the states $N(1440)$, $N(1520)$, 
and $N(1535)$, provided that it is 
allowed to release energy to turn itself  into $p\bar p$, and the mass relation of $m_X\simeq m_{N^*}+m_{\bar N^*}$
can be simply satisfied. 
Note that $\Lambda_c(2800)$ and $\Lambda_c(2940)$ as excited charmed baryon states
are proposed to be
$DN$ and $D^* p$ bound states~\cite{Dong, XGHe}, respectively.
However, at present, it is impossible for us
 to distinguish whether the resonance is the bound state or the glueball state
as they carry the same quantum numbers~\cite{chunLiu}.

In sum, we have identified the existence of  the glueball state at 3.02 GeV based on
 the peak in the $m_{p\bar p}$ spectrum of  $\bar B^0\to p\bar p D$ for $m_{D p}<3$ GeV
observed by the BABAR collaboration, which could be the first glueball state
 above 3 GeV. Explicitly, it has been fitted to be $X(3020)$
 with $(m_X,\;\Gamma_X)=(3020\pm 8,\; 107\pm 30)\;\text{MeV}$.

\begin{acknowledgments}
The work was supported in part by National Center for Theoretical Sciences,  National Science
Council (NSC-98-2112-M-007-008-MY3 and NSC-101-2112-M-007-006-MY3) and
National Tsing-Hua University (102N2725E1),  Taiwan, R.O.C.
\end{acknowledgments}


\begin{thebibliography}{99}
\bibitem{derivation} 
The reduced constants are determined in Ref. \cite{ppD} to be
$(D_{||},\;D_{\overline{||}})=(67.7\pm 16.3,\,-280.0\pm 35.9)\;{\rm GeV^5}$ and
$(D_{||}^2,\,D_{||}^3,\,D_{||}^4,\,D_{||}^5)
=(-187.3\pm 26.6,\,-840.1\pm 132.1,\,-10.1\pm 10.8,\,-157.0\pm 27.1)\;{\rm GeV^4}$.
\end{thebibliography}
\end{document}